# A verified equivalent-circuit model for slot-waveguide modulators

Heiner Zwickel,[1] Stefan Singer,[1] Clemens Kieninger,[1,2] Yasar Kutuvantavida,[1,2] Narek Muradyan,[1] Thorsten Wahlbrink,[3] Shiyoshi Yokoyama,[4] Sebastian Randel,[1] Wolfgang Freude,[1] and Christian Koos[1,2,*]

[1]Institute of Photonics and Quantum Electronics (IPQ), Karlsruhe Institute of Technology (KIT), Germany
[2]Institute of Microstructure Technology (IMT), Karlsruhe Institute of Technology (KIT), Germany
[3]AMO GmbH, Aachen, Germany
[4]Institute for Materials Chemistry and Engineering, Kyushu University, Japan
*christian.koos@kit.edu

**Abstract:** We formulate and experimentally validate an equivalent-circuit model based on distributed elements to describe the electric and electro-optic (EO) properties of travelling-wave silicon-organic hybrid (SOH) slot-waveguide modulators. The model allows to reliably predict the small-signal EO frequency response of the modulators exploiting purely electrical measurements of the frequency-dependent RF transmission characteristics. We experimentally verify the validity of our model, and we formulate design guidelines for an optimum trade-off between optical loss due to free-carrier absorption (FCA), electro-optic bandwidth, and π-voltage of SOH slot-waveguide modulators.

## 1. Introduction

Broadband efficient electro-optic (EO) modulators are key elements for a wide variety of applications, ranging from high-speed optical communications [1–6] to ultra-fast signal processing [7,8] and further to optical metrology and sensing [9,10]. Among the various technology platforms that are available today, silicon photonics (SiP) represents a particularly promising option, exploiting mature CMOS fabrication processes for cost-efficient mass production of densely integrated photonic circuits [11–13]. When it comes to EO modulators, hybrid combinations of SiP slot waveguides with highly efficient organic EO cladding materials are an attractive approach to overcome the intrinsic lack of second-order optical nonlinearities in bulk silicon [14]. Exploiting theory-guided optimization of organic EO materials on a molecular level [15], these silicon-organic hybrid (SOH) devices primarily stand out due to small voltage-length products down to 0.32 Vmm [16] — more than an order of magnitude below that of conventional depletion-type SiP modulators [17]. SOH devices have been demonstrated to offer attractive performance parameters such as EO bandwidths of 100 GHz [18] as well as line rates of up to 120 Gbit/s for intensity modulation [5,19] and up to 400 Gbit/s for coherent 16QAM signaling [6], while supporting driver-less operation with sub-1 V drive signals obtained from binary outputs of a standard CMOS FPGA [20]. However, the EO bandwidth of these devices is still subject to large uncertainties. In addition, most of the aforementioned experiments [5,6,18,19,21] relied on a so-called gate voltage that is applied across the buried oxide (BOX) of the underlying silicon-on-insulator (SOI) substrate to induce a highly conductive electron accumulation layer that reduces the RC time constant of the slot waveguide [22]. Moreover, an experimentally verified model describing the EO frequency response of SOH modulators is still lacking, which represents a major obstacle towards a targeted design of the modulator structures and of the associated drive circuitry.

In this paper, we formulate and experimentally verify a reliable quantitative model of slot-waveguide modulators that accounts for the various bandwidth-limiting effects. We build upon an equivalent-circuit approach proposed by Witzens *et al.* [23], which relies on a distributed-element description of the slot waveguide structure and of the associated radio-frequency (RF) transmission line. We refine this model and verify it quantitatively using experimental data obtained from SOH devices. We find that the model accurately predicts the EO frequency response of a slot-waveguide modulator based on purely electrical RF characteristics of the device, which are easily accessible by measurements. Our model allows to identify non-optimum doping profiles as the main reason for a limited EO bandwidth of our current device generation. Based on these findings, we extract design rules that lead to an optimum trade-off between optical loss due to free-carrier absorption (FCA), EO bandwidth, and π-voltage. Optimized device designs should lead to sub-1 mm Mach-Zehnder modulators (MZM) with π-voltages smaller than 1 V, electro-optic-electric (EOE) bandwidths larger than 100 GHz, and FCA-induced optical losses below 0.1 dB. To the best of our knowledge, this work represents the first experimental validation of a general model of slot-waveguide modulators. We believe that quantitatively reliable models represent a key step towards fast and reliable design and efficient wafer-level characterization of slot-waveguide modulators which are embedded into complex optical systems and RF drive circuits.

The paper is structured as follows: In Section 2, we introduce the concept of SOH modulators and formulate a model based on distributed elements. In Section 3, we discuss *S*-parameter measurements of SOH modulators, extract the characteristic impedance and the RF propagation parameter, and determine the parameters of the equivalent RF circuit by a least-squares fit. In Section 4, we validate the model by comparing the measured EO frequency response to the model prediction using parameters extracted from purely electrical measurements. In Section 5, we summarize the findings and discuss design guidelines and achievable performance parameters of SOH EO modulators.

## 2. Device principle and model based on distributed-element equivalent circuit

The concept of a slot-waveguide SOH Mach-Zehnder modulator (MZM) is depicted in Figure 1 [24–26]. The MZM consists of two parallel phase shifter sections having a length $\ell$ of typically 0.5 mm ... 1.5 mm, see Fig. 1(a). The electric drive signal is applied via a coplanar RF transmission line in ground-signal-ground (GSG) configuration. The device is realized as a travelling-wave structure, in which the modulating electric mode co-propagates with the optical mode. Each phase shifter comprises a silicon slot waveguide with a slot width of typically 60 ... 160 nm, which is filled with a highly efficient organic electro-optic (EO) material, that may offer in-device EO coefficients in excess of 300 pm/V [16], see cross-section in Fig. 1(b). To activate the macroscopic EO activity in the slot, chromophores in the organic EO material have to be aligned in a one-time poling process, see [24,27] for details. The rails of the slot waveguide are connected to the metal RF transmission line via thin, doped silicon slabs, such that the externally applied drive voltage predominantly drops across the narrow slot. At the same time, the optical mode is highly confined to the slot region, leading to a very strong interaction with the modulating electric field. The EO frequency response of SOH modulators strongly depends on the RC time constant that is associated with the slot capacitance and the slab resistance. A two-level doping profile as illustrated in Fig. 1(b) can help to increase the bandwidth while maintaining acceptable optical loss [28,29]: In the rails and near the slot, the doping concentration is kept low (light blue) to avoid excessive free-carrier absorption, while a higher doping concentration (dark blue) is used further away from the optical waveguide to increase the electric conductivity. The impact of increased slab conductivity can also be studied by applying a DC "gate" voltage $U_{gate}$ between the substrate and the slabs. This induces a highly conductive electron accumulation layer at the interface between the Si slabs and the BOX [22].

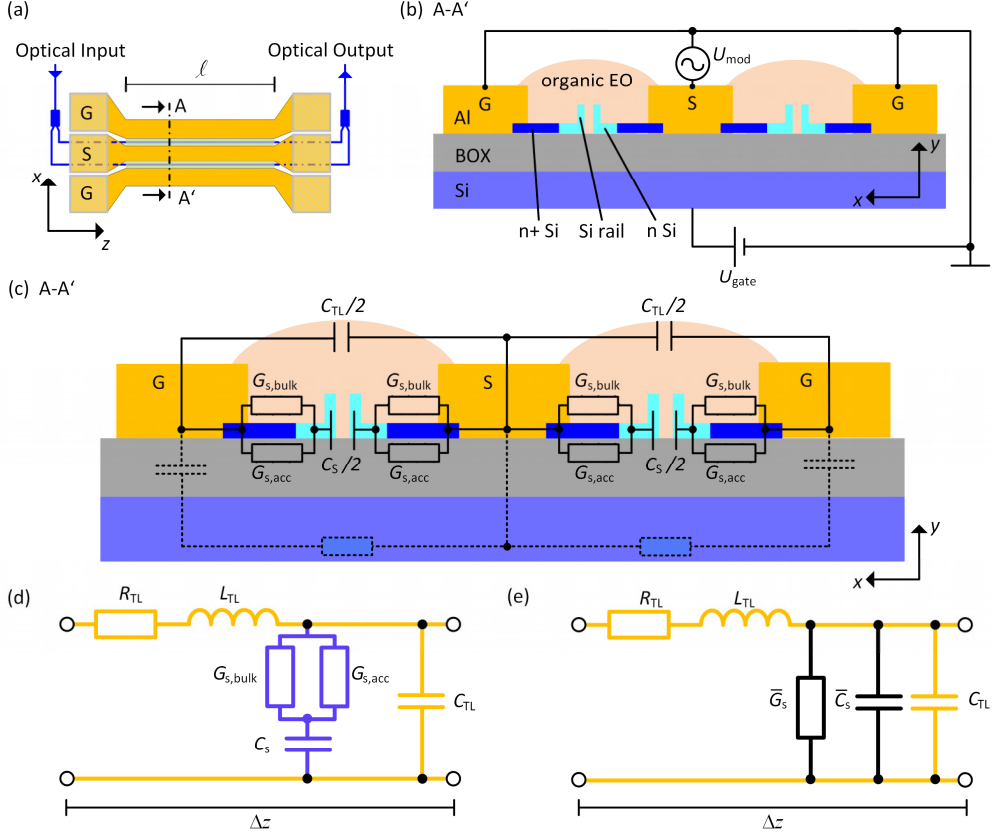

Fig. 1. Device concept and equivalent-circuit model. **(a)** Schematic top view of the Mach-Zehnder modulator (MZM) with metal electrodes (yellow) that form a coplanar ground-signal-ground electrical transmission line of length $\ell$. The device is equipped with contact pads at both ends. The optical waveguide depicted in blue is split into two arms to realize the Mach-Zehnder interferometer. **(b)** Cross-section of the MZM along the line A-A' indicated in Subfigure (a). Each arm comprises a SiP slot waveguide filled with an organic electro-optic (EO) material. The metal electrodes are connected to the rails of the slot waveguide via thin, doped Si slabs. A modulating voltage $U_{\mathrm{mod}}$ applied to the signal electrode, drops mainly across the narrow slot. This leads to a strong electric field in the slot that changes the refractive index of the organic EO material and thus modulates the phase of the optical wave. The Si slabs are doped to increase the conductivity. A low doping concentration (light blue) is used near the slot to keep optical losses low. A higher doping concentration (dark blue) is used further away from the slot to achieve a low resistance of the slabs and to thus decrease the RC time constant of the modulator. In addition, a DC "gate" voltage $U_{\mathrm{gate}}$ can be applied between the substrate and the ground electrodes. This induces a highly conductive electron accumulation layer at the interface between the Si slabs and the buried oxide layer (BOX) and thus further decreases the RC time constant. **(c)** Enlarged cross-sectional view of the MZM along with the associated elements of an equivalent-circuit representation of the modulator. The capacitance of the slot is represented by a capacitor with capacitance $C_{\mathrm{S}}/2$. The finite conductivity of each slab is represented by two parallel conductances $G_{\mathrm{S,bulk}}$ and $G_{\mathrm{S,acc}}$ that account for the conductance of the bulk Si slab and of the accumulation layer formed by the gate voltage, respectively. The metal electrodes have a total capacitance $C_{\mathrm{TL}}$. **(d)** Equivalent-circuit model of an infinitesimally short element of the electrical transmission line formed by the MZM. The distributed elements shown in yellow correspond to the metal electrodes. The distributed elements shown in blue represent the additional RC shunt load formed by the slot waveguide. **(e)** Telegrapher-type equivalent-circuit representation of an MZM transmission-line element [23]. This representation is obtained by transforming the circuit in Subfigure (d), using a frequency-dependent shunt conductance $\overline{G}_{\mathrm{S}}$ and a frequency-dependent shunt capacitance $\overline{C}_{\mathrm{S}}$, see Eq. (2) and Eq. (3).

For a quantitative description of the electric behavior of the SOH modulator, we adapt the model suggested in [23], which has also been adapted to model depletion-type pn-modulators [30–36]. Reference [23] is primarily focused on the design of phase-matched travelling-wave transmission lines, which is key to achieve broadband operation of MZM with rather large interaction lengths $\ell$ of several millimeters. Over the previous years, however, substantial improvements of organic EO materials and in-device EO efficiencies [16] have allowed to significantly reduce the interaction length. This renders phase-matching less crucial and requires adapted design considerations that focus on the RC time constant of the slab/slot configuration and on the associated trade-off with respect to optical loss.

In our analysis, we rely on a distributed-element model of the underlying RF transmission line, see Fig. 1(c) for a cross-sectional view with elements of an equivalent circuit representing an infinitesimally short transmission-line section of length $\Delta z$. The full equivalent circuit is shown in Fig. 1(d). First, we consider the circuit elements depicted in yellow in Fig. 1(d) that describe the coplanar GSG transmission line formed by the metal electrodes only, without the optical slot-waveguide structure. These circuits elements comprise a series resistance $R'_{\mathrm{TL}}\Delta z$, a series inductance $L'_{\mathrm{TL}}\Delta z$, and a shunt capacitance $C'_{\mathrm{TL}}\Delta z$, where $R'_{\mathrm{TL}}$, $L'_{\mathrm{TL}}$, and $C'_{\mathrm{TL}}$ are differential quantities that describe the resistance, inductance, and capacitance of the GSG transmission line per length in $\Omega$/m, H/m, and F/m, respectively. This corresponds to the classical telegrapher circuit with negligible shunt conductance. In a second step, we expand the equivalent circuit by the electrical representation of the Si structure (blue) forming the optical slot waveguide. For practical doping concentrations, the conductivity of the slabs is much smaller than that of the metal electrodes. As a consequence, the longitudinal currents can be assumed to be confined to the metal electrodes, whereas the currents in the silicon slabs flow predominantly in the transverse (*x*) direction between the electrodes and the slot. These assumptions are in accordance with findings from finite-element simulations [23] and allow us to model the electrical behavior of the optical slot waveguide by simply adding an RC shunt to the equivalent circuit [23,30,37]. The associated elements are the differential capacitance $C'_{\mathrm{S}}$ of the pair of slots, the differential conductance $G'_{\mathrm{S,bulk}}$ of the resistive Si slabs to both sides of each of the slot waveguides, and the differential conductance $G'_{\mathrm{S,acc}} = g'_{\mathrm{S,acc}} U_{\mathrm{gate}}$, which accounts for the increased conductance of an accumulation layer that may be induced by applying a gate voltage $U_{\mathrm{gate}}$. Note that the capacitance between center signal electrode and each of the ground electrodes amounts to $C_{\mathrm{TL}}/2$, and that the overall shunt capacitance of the metal electrodes forming the coplanar GSG transmission line is hence $C_{\mathrm{TL}} = C'_{\mathrm{TL}}\Delta z$. Similarly, $C_{\mathrm{S}}/2$ is the capacitance of an individual slot, and $C_{\mathrm{S}} = C'_{\mathrm{S}}\Delta z$ represents the combination of both slots. The quantities $G_{\mathrm{S,bulk}} = G'_{\mathrm{S,bulk}}\Delta z$ and $G_{\mathrm{S,acc}} = G'_{\mathrm{S,acc}}\Delta z$ refer to the slab conductance to either side of each slot, and the equivalent circuit-element in Fig. 1(d) represents the combined effect of four slabs. The total differential conductance is $G'_{\mathrm{S}} = G'_{\mathrm{S,bulk}} + G'_{\mathrm{S,acc}}$. The two slot waveguides thus create a differential shunt admittance $\underline{Y}'_S$ given by

$$\frac{1}{\underline{Y}'_S} = \frac{1}{G'_{\mathrm{S,bulk}} + G'_{\mathrm{S,acc}}} + \frac{1}{j\omega_{\mathrm{RF}} C'_{\mathrm{S}}}. \tag{1}$$

For completeness, we have also indicated additional elements by dashed lines in Fig. 1(c) that seem to be reasonable to include. These elements model an additional RC shunt that accounts for the finite resistance of the handle wafer and the capacitance created by the buried oxide (BOX) layer. It turned out, however, that in our case the impact of this additional RC circuit is negligible despite the rather low substrate resistivity of about 20 Ωcm as specified by the wafer manufacturer. This additional RC shunt load is therefore disregarded in the following, but could be easily included if modified device or material choices would require an adapted model [30].

The overall circuit shown in Fig. 1(d) can be transformed into the circuit in Fig. 1(e) by introducing the frequency-dependent differential conductance

$$\overline{G}'_S(\omega_{RF}) = \Re\{\underline{Y}'_S\} \tag{2}$$

and the frequency-dependent differential capacitance

$$\overline{C}'_S(\omega_{RF}) = \frac{1}{\omega_{RF}} \Im\{\underline{Y}'_S\}. \tag{3}$$

This circuit resembles that of a classical telegrapher-type transmission-line model, and the transmission line can thus be fully described by the complex propagation parameter $\underline{\gamma} = \alpha + j\beta$ together with the complex characteristic impedance $\underline{Z}_0$ obtained from the well-known transmission-line equations [38]

$$\underline{\gamma} = \alpha + j\beta = \sqrt{(R'_{TL} + j\omega_{RF}L'_{TL})(\overline{G}'_S + j\omega_{RF}(\overline{C}'_S + C'_{TL}))}, \tag{4}$$

$$\underline{Z}_0 = \sqrt{\frac{R'_{TL} + j\omega_{RF}L'_{TL}}{\overline{G}'_S + j\omega_{RF}(\overline{C}'_S + C'_{TL})}}. \tag{5}$$

Note that in these relations $\alpha$ denotes an amplitude attenuation coefficient. The corresponding power attenuation coefficient amounts to $2\alpha$.

For analyzing the EO interaction, we now consider a travelling-wave EO modulator with characteristic impedance $\underline{Z}_0$, complex RF propagation parameter $\underline{\gamma}$, optical group refractive index $n_{g,opt}$ in the phase shifter, and active phase-shifter length $\ell$. The device is connected to an RF source with impedance $\underline{Z}_{src}$ and terminated by an impedance $\underline{Z}_{trm}$. Due to the Pockels effect, the local optical phase shift increases in proportion to the local modulation voltage. If a sinusoidal voltage with angular frequency $\omega_{RF}$ is applied to the modulator's signal electrode, the total phase shift $\varphi(t) = \Re\{\hat{\varphi}(\omega_{RF})e^{j\omega_{RF}t}\}$ in each arm can then be obtained from an integral of the modulation voltage seen by the optical signal along the phase shifter length $\ell$, see ref. [39], Eq. (2). In this relation, the complex amplitude $\hat{\varphi}(\omega_{RF})$ contains both the frequency-dependent amplitude and the phase of the sinusoidally varying phase shift $\varphi(t)$. The complex frequency response $\underline{m}_{PH}(\omega_{RF})$ of such a phase modulator is given by the phase amplitude $\hat{\varphi}(\omega_{RF})$, normalized to its value that would be obtained for an angular frequency $\omega_{RF} = 0$ for a source and a termination impedance that are both perfectly matched to the modulator's characteristic transmission-line impedance $\underline{Z}_0$ ([39], Eq. (3)). This leads to

$$\underline{m}_{PH}(\omega_{RF}) = \frac{\hat{\varphi}(\omega_{RF})}{\hat{\varphi}(0)|_{\underline{Z}_{src}=\underline{Z}_{trm}=\underline{Z}_0}} = \underline{A}\,\underline{H}_{RC}\left(\frac{e^{\underline{q}^-\ell}-1}{\underline{q}^-} + \underline{B}\frac{e^{\underline{q}^+\ell}-1}{\underline{q}^+}\right), \tag{6}$$

where

$$\underline{A} = \frac{2}{\ell}\frac{\underline{Z}_0}{\underline{Z}_0 + \underline{Z}_{src}}\frac{1}{1-\underline{\Gamma}_{src}\underline{\Gamma}_{trm}\exp(-2\underline{\gamma}\ell)}, \tag{7}$$

$$\underline{B} = \exp(-2\underline{\gamma}\ell)\underline{\Gamma}_{trm}, \tag{8}$$

$$\underline{q}^{\pm} = \pm \underline{\gamma} + \mathrm{j}\, n_{\mathrm{g,opt}} \omega_{\mathrm{RF}}/c, \tag{9}$$

and where the reflection factors at the source and the termination are given by

$$\underline{\Gamma}_{\mathrm{src}} = \frac{\underline{Z}_{\mathrm{src}} - \underline{Z}_0}{\underline{Z}_{\mathrm{src}} + \underline{Z}_0} \quad \text{and} \quad \underline{\Gamma}_{\mathrm{trm}} = \frac{\underline{Z}_{\mathrm{trm}} - \underline{Z}_0}{\underline{Z}_{\mathrm{trm}} + \underline{Z}_0}. \tag{10}$$

The quantity $\underline{H}_{\mathrm{RC}}(\omega_{\mathrm{RF}})$ in Eq. (6) accounts for the fact that the voltage that becomes effective for modulation in the slot waveguide is reduced with respect to the voltage on the metal transmission lines through a voltage divider formed by the slab conductance and the slot capacitance [30]. This reduction becomes more pronounced as the frequency increases and can be written as

$$\underline{H}_{\mathrm{RC}}(\omega_{\mathrm{RF}}) = \frac{1}{1 + \mathrm{j}\omega_{\mathrm{RF}} C'_{\mathrm{S}} / G'_{\mathrm{S}}}. \tag{11}$$

The frequency response of EO modulators is usually obtained from a frequency-dependent decay of an intensity modulation rather than from a phase modulation. To connect the two quantities, we consider an MZM in push-pull configuration, for which the phase shifts $\varphi_1$ and $\varphi_2$ in the two arms fulfill the relation $\varphi_1 = -\varphi_2 = \varphi$. The phase difference between the two arms $\Delta\varphi$ is then given by $\Delta\varphi = 2\varphi_1$, and the intensity $I_{\mathrm{opt}}$ of the optical output signal is related to the phase difference $\Delta\varphi$ by

$$I_{\mathrm{opt}} \propto \cos^2(\Delta\varphi/2) = \cos^2(\varphi_1). \tag{12}$$

If the push-pull MZM is biased at the 3 dB point (quadrature point) and operated under small-signal conditions, i.e., with $\Delta\varphi = \pi/2 \pm \delta\varphi$ ($|\delta\varphi| \ll \pi/2$), Eq. (12) can be linearized and the change in optical power $\delta P_{\mathrm{opt}}$ is in proportion to $\delta\varphi$. In characterization experiments, the optical output of the modulator is usually coupled to a photodetector, the photocurrent $I_{\mathrm{p}}$ of which is proportional to the received optical power such that

$$\delta I_{\mathrm{p}} \propto \delta P_{\mathrm{opt}} \propto \delta\varphi. \tag{13}$$

The small-signal intensity frequency response $\underline{m}_{\mathrm{EOE}}(\omega_{\mathrm{RF}})$ of the electro-optic-electric (EOE) conversion is therefore identical to the phase modulator frequency response given in Eq. (6),

$$\underline{m}_{\mathrm{EOE}}(\omega_{\mathrm{RF}}) = \underline{m}_{\mathrm{PH}}(\omega_{\mathrm{RF}}). \tag{14}$$

To quantify the bandwidth of the modulator, the 3 dB corner frequency $f_{\mathrm{3dB,EOE}} = \omega_{\mathrm{3dB,EOE}}/2\pi$ of the intensity modulation may be used. The 3 dB corner frequency $\omega_{\mathrm{3dB,EOE}}/2\pi$ is defined as the frequency $\omega_{\mathrm{RF}}/2\pi$, for which $|\underline{m}_{\mathrm{EOE}}(\omega_{\mathrm{RF}})|^2$ is reduced by a factor of two compared to $|\underline{m}_{\mathrm{EOE}}(0)|^2$, which corresponds to a power decay of the associated spectral component of the photocurrent by a factor of two. Alternatively, also the 6 dB corner frequency $f_{\mathrm{6dB,EOE}} = \omega_{\mathrm{6dB,EOE}}/2\pi$ of $|\underline{m}_{\mathrm{EOE}}(\omega_{\mathrm{RF}})|^2$ can be considered, which corresponds to a decay of $|\underline{m}_{\mathrm{EOE}}(\omega_{\mathrm{RF}})|^2$ and of the associated photocurrent power by a factor of four and is hence equivalent to a reduction of the photocurrent amplitude or, equivalently, the phase modulation amplitude by a factor of two. Both values are used in the literature to refer to the bandwidth of a modulator [40–42], sometimes without explicitly stating which of the two values is specified. In the following, we either specify both values or reside to the 6 dB corner frequency for better comparability to similar investigations of SiP EO modulators [23,26,30,42].

To apply the model developed here to a given device design, the complex propagation parameter $\underline{\gamma}$ and the characteristic impedance $\underline{Z}_0$ can be directly determined from S-parameter measurements or with numerical mode solvers. The equivalent-circuit parameters $R'_{\mathrm{TL}}, L'_{\mathrm{TL}}, C'_{\mathrm{TL}}, G'_{\mathrm{S}}, C'_{\mathrm{S}}$ can then be obtained by fitting Eq. (4) and Eq. (5) to the frequency-dependence of $\underline{\gamma}$ and $\underline{Z}_0$, see Section 3. If, in addition, $n_{\mathrm{g,opt}}$ is known, Eq. (6) allows to predict the EO response of the modulator.

## 3. Electrical characterization and fitting procedure

To experimentally verify the model introduced in Section 2, we determine the frequency-dependent characteristic impedance $\underline{Z}_0$ and propagation parameter $\underline{\gamma}$ of a typical SOH modulator. These parameters fully describe the electrical behavior of the device and build the base for determining the frequency response $|m_{\mathrm{EOE}}(\omega_{\mathrm{RF}})|$ of the electro-optic-electric (EOE) conversion, see Section 4. For the devices used here, the thickness of the BOX layer was 3 µm, the device layer thickness was 220 nm, and the slot width was 120 nm. To de-embed the characteristics of the devices-under-test (DUT) from those of the RF contact-pad parasitics, we investigate two MZM with nominally identical cross sections and different active lengths $\ell_1 = 500\,\mu\mathrm{m}$ and $\ell_2 = 750\,\mu\mathrm{m}$ of the phase shifters. In a first step, we measure S-parameters of the embedded devices using a vector network analyzer (VNA) and microwave contact probes. Figure 2(c) shows the measured (embedded) transmission $\underline{S}_{21}$ and reflection $\underline{S}_{11}$ of the two modulators for a gate voltage $U_{\mathrm{gate}} = 0$ V. The reference plane was moved to the tip of the probes using a short-open-load-through (SOLT) calibration routine with an impedance-standard substrate. The phase-shifter sections of both devices are thus only embedded into identical on-chip contact circuits that are formed by the contact pads and subsequent transitions on both ends of the modulator, see Fig. 2(a), and that are referred to as contact fixtures in the following. Each of these fixtures is modeled by a series impedance $\underline{Z}$ and a shunt admittance $\underline{Y}$ as shown in Fig. 2(b), which can be represented by electrical transfer matrices (T-matrices [38]) $\underline{T}_{\mathrm{L}}$ and $\underline{T}_{\mathrm{R}}$ for the left and the right fixture, respectively, see Appendix A, Eq. (A1) and Eq. (A2). These T-matrices can be calculated from the measured S-parameters $\underline{S}_{21}$ and $\underline{S}_{11}$ using standard microwave theory, see Appendix A. To de-embed the T-matrices $\underline{T}_{\mathrm{DUT1}}$ and $\underline{T}_{\mathrm{DUT2}}$ of the phase shifters with lengths $\ell_1$ and $\ell_2$, we first extract the measured T-matrix of the embedded modulator $\underline{T}_{\mathrm{m1}} = \underline{T}_{\mathrm{L}}\underline{T}_{\mathrm{DUT1}}\underline{T}_{\mathrm{R}}$ and $\underline{T}_{\mathrm{m2}} = \underline{T}_{\mathrm{L}}\underline{T}_{\mathrm{DUT2}}\underline{T}_{\mathrm{R}}$ from the measured S-parameters. Introducing the T-matrix $\underline{T}_{\mathrm{DUT0}}$ of an $\ell_0 = 250\,\mu\mathrm{m}$-long de-embedded modulator section and exploiting that $\ell_1 = 2\ell_0$ and $\ell_2 = 3\ell_0$, we can rewrite $\underline{T}_{\mathrm{DUT1}}$ and $\underline{T}_{\mathrm{DUT2}}$ as $\underline{T}_{\mathrm{DUT1}} = \underline{T}_{\mathrm{DUT0}}\underline{T}_{\mathrm{DUT0}}$ and $\underline{T}_{\mathrm{DUT2}} = \underline{T}_{\mathrm{DUT0}}\underline{T}_{\mathrm{DUT0}}\underline{T}_{\mathrm{DUT0}}$. The T-matrix of the virtual through connection shown in Fig. 2(b) is defined as $\underline{T}_{\mathrm{thru}} = \underline{T}_{\mathrm{L}}\underline{T}_{\mathrm{R}}$ and can then be calculated from the measured matrices as $\underline{T}_{\mathrm{thru}} = \underline{T}_{\mathrm{m1}}\underline{T}_{\mathrm{m2}}^{-1}\underline{T}_{\mathrm{m1}}\underline{T}_{\mathrm{m2}}^{-1}\underline{T}_{\mathrm{m1}}$. This allows to determine $\underline{T}_{\mathrm{L}}$ and $\underline{T}_{\mathrm{R}}$ directly from the measured transfer matrices $\underline{T}_{\mathrm{m1}}$ and $\underline{T}_{\mathrm{m2}}$ and to de-embed the modulator response by calculating $\underline{T}_{\mathrm{DUT1}} = \underline{T}_{\mathrm{L}}^{-1}\underline{T}_{\mathrm{m1}}\underline{T}_{\mathrm{R}}^{-1}$ and $\underline{T}_{\mathrm{DUT2}} = \underline{T}_{\mathrm{L}}^{-1}\underline{T}_{\mathrm{m2}}\underline{T}_{\mathrm{R}}^{-1}$. Using Eq. (A7) of Appendix A, we can then extract the frequency-dependent characteristic impedance $\underline{Z}_0$ and the frequency-dependent propagation parameter $\underline{\gamma}$ from the transfer matrices $\underline{T}_{\mathrm{DUT1}}$ and $\underline{T}_{\mathrm{DUT2}}$ of the de-embedded modulator, see black traces in Figures 2(d-g). We find that the results obtained from $\underline{T}_{\mathrm{DUT1}}$ and $\underline{T}_{\mathrm{DUT2}}$ show very good agreement, thereby confirming the validity of our de-embedding approach. For comparison, we also extract $\underline{Z}_0$ and $\underline{\gamma}$ by applying Eqs. (A3-A7) directly to the measured T-matrices $\underline{T}_{\mathrm{m1}}$ and $\underline{T}_{\mathrm{m2}}$ of the embedded modulators, thereby ignoring the impact of the contact fixtures. The resulting frequency characteristics of $\underline{Z}_0$ and $\underline{\gamma}$ are depicted as blue and red traces in Figures 2(d-g) and show clear differences, thereby underlining the importance of proper de-embedding.

Our results indicated that $\underline{Z}_0$ has a negligible imaginary part and amounts to approximately 50 Ω – except for the low-frequency region, where $\underline{Z}_0$ deviates from its ideal low-loss and high-frequency approximation, see Eq. (B1) in Appendix B. This ensures good impedance matching to typical feeding networks and drivers. The real part $\alpha$ of $\underline{\gamma}$, associated with RF propagation loss, increases with frequency due to loss in the metal stripes and in the resistive slabs, see the following paragraphs for a detailed discussion of the loss mechanisms. The imaginary part $\beta$ of $\underline{\gamma}$ shows a linear increase indicating a low dispersion of the RF transmission line.

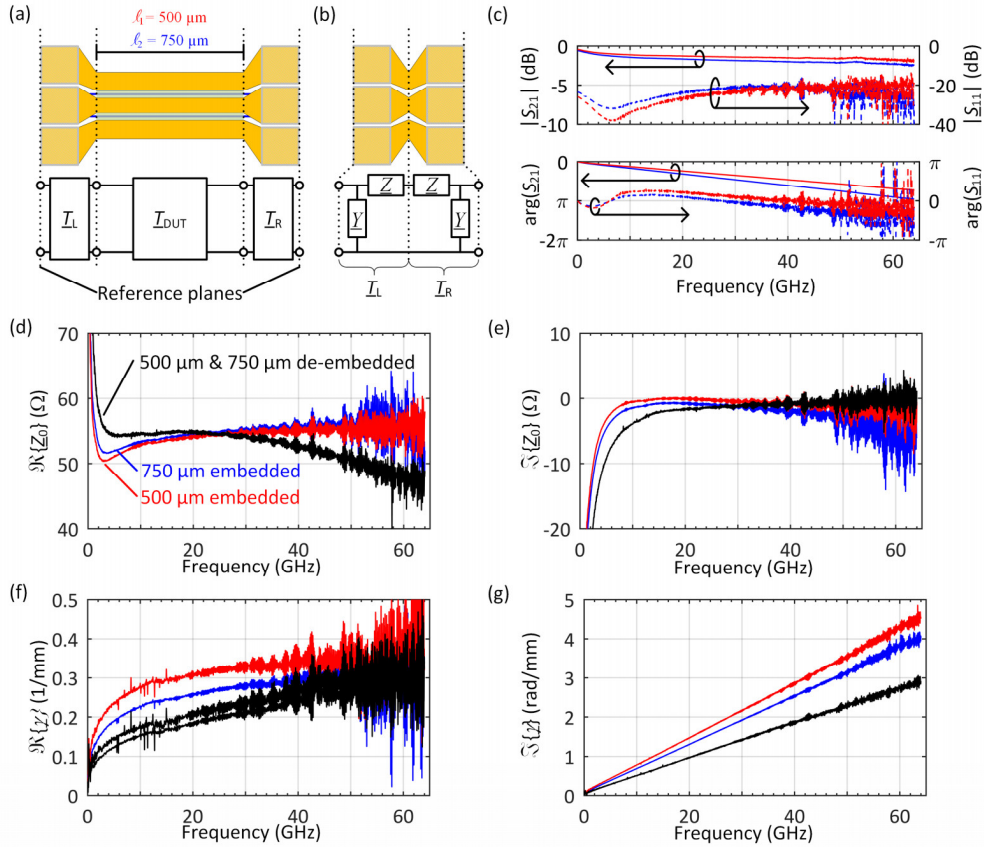

Fig. 2. **(a)** Simplified electrical structure of the SOH modulators. The phase-shifter sections with lengths $\ell_1$ or $\ell_2$ and $T$-matrices $\underline{T}_{DUT1}$ or $\underline{T}_{DUT2}$ are electrically embedded into contact fixtures, modeled by $T$-matrices $\underline{T}_L$ and $\underline{T}_R$. **(b)** The contact fixtures are modelled by a series impedance $\underline{Z}$ and a shunt admittance $\underline{Y}$. **(c)** Amplitude and phase of measured $S$-parameters for embedded MZM with lengths $\ell_1 = 500\,\mu m$ (red) and $\ell_2 = 750\,\mu m$ (blue). **(d)** Real part and **(e)** imaginary part of the characteristic impedance $\underline{Z}_0$, along with **(f)** real part and **(g)** imaginary part of the propagation constant $\underline{\gamma}$. Black curves indicate values as extracted from de-embedded modulators of lengths $\ell_1$ and $\ell_2$. These results show good agreement and hence confirm the validity of our de-embedding approach. In contrast to that, the red and blue curves correspond to values extracted from the embedded modulators by ignoring the impact of the contact fixtures. Consequently, the results obtained for the different device lengths are not in agreement.

The measurement of $S$-parameters and subsequent de-embedding and extraction of $\underline{\gamma}$ and $\underline{Z}_0$ was repeated for a range of gate voltages $U_{\text{gate}} = 0\ldots 300$ V. Figure 3 shows the numerical values obtained from measurements in blue. To determine the six equivalent-circuit parameters $R'_{\text{TL}}, L'_{\text{TL}}, C'_{\text{TL}}, G'_{\text{S,bulk}}, G'_{\text{S,acc}}$, and $C'_{\text{S}}$, the following approach was used: First, Eq. (4) and Eq. (5) were used to calculate $R'_{\text{TL}} = \Re\{\underline{Z}_0 \underline{\gamma}\}$ and $L'_{\text{TL}} = \Im\{\underline{Z}_0 \underline{\gamma}\}/\omega_{\text{RF}}$ directly from $\underline{\gamma}$ and $\underline{Z}_0$. Using the fact that both $R'_{\text{TL}}$ and $L'_{\text{TL}}$ are independent of frequency and of the applied gate voltage, the two values were calculated for all frequencies and gate voltages and then averaged. In a second step, a multi-dimensional multi-parameter least-squares fit was used to obtain $C'_{\text{TL}}, G'_{\text{S,bulk}}, G'_{\text{S,acc}}$ and $C'_{\text{S}}$ by fitting Eq. (4) and Eq. (5) to the measured frequency-dependent and gate-voltage-dependent values of $\underline{\gamma}$ and $\underline{Z}_0$, see Table 1 for the resulting numerical values. The fitted curves are shown in red in Fig. 3 and show good agreement with the measured data. For increasing frequencies, the characteristic impedance quickly approaches its ideal high-frequency real value $\underline{Z}_0 = \sqrt{L'_{TL}/(\overline{C}'_{\text{S}} + C'_{\text{TL}})} \approx 50\,\Omega$, which can be obtained from Eq. (5) using $R'_{\text{TL}} \ll \omega_{\text{RF}} L'_{\text{TL}}$ and $\overline{G}'_{\text{S}} \ll \omega_{\text{RF}}(\overline{C}'_{\text{S}} + C'_{TL})$. The characteristic impedance $\underline{Z}_0$ as well as the imaginary part of $\underline{\gamma}$ are rather insensitive to the gate voltage.

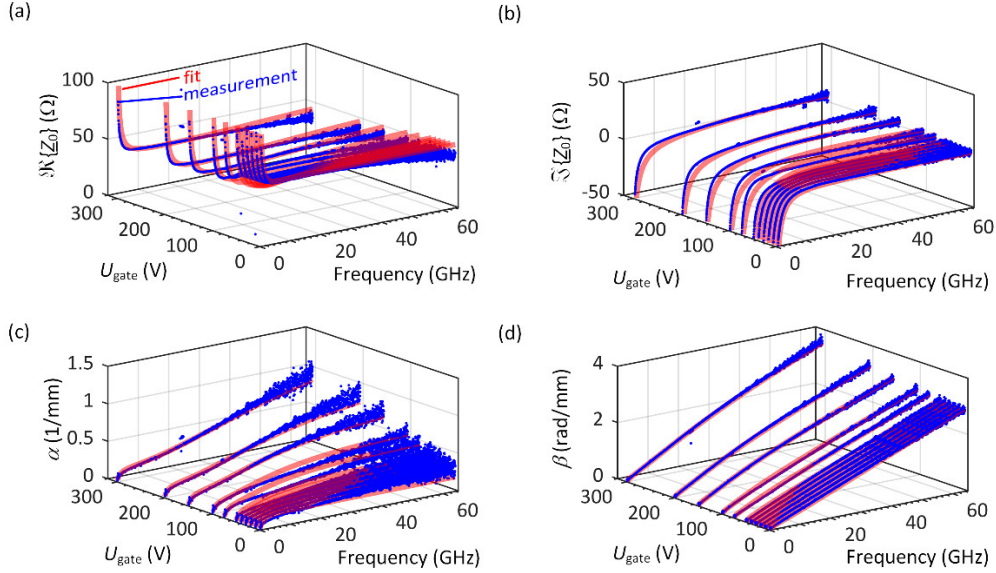

Fig. 3. Measured values (blue) and fitted curves (red) according to Eq. (4) and (5) as a function of both RF frequency and gate voltage $U_{\text{gate}}$ for **(a)** the real part of the characteristic impedance $\underline{Z}_0$, **(b)** the imaginary part of the characteristic impedance $\underline{Z}_0$, **(c)** the real part of the complex propagation parameter $\underline{\gamma}$ associated with loss, and **(d)** the imaginary part of the complex propagation parameter $\underline{\gamma}$.

**Table 1 Circuit parameters obtained by fitting Eq. (4) and (5) to the measured frequency-dependent and gate-voltage-dependent values of $\underline{\gamma}$ and $\underline{Z}_0$, see Fig. 3.**

|  | Fit |  |
| --- | --- | --- |
| $R'_{\text{TL}}$ | 13 | Ω/mm |
| $L'_{\text{TL}}$ | 414 | pH/mm |
| $C'_{\text{TL}}$ | 130 | fF/mm |
| $G'_{\text{S,bulk}}$ | 2.76 | mS/mm |
| $g'_{\text{S,acc}} = G'_{\text{S,acc}}/U_{\text{gate}}$ | 145 | μS/(Vmm) |
| $C'_{\text{S}}$ | 160 | fF/mm |

In contrast to this, the losses show a strong dependence on the conductivity of the slabs and thus on $U_{\text{gate}}$. To explain this finding, we first consider the losses in the metal transmission line represented by $R'_{\text{TL}}, L'_{\text{TL}}$ and $C'_{\text{TL}}$ separately from the losses in the RC element which represents the slab and the slot and which is formed by $G'_{S,\text{bulk}}, G'_{S,\text{acc}}$ and $C'_S$. The different loss contributions are plotted as a function of frequency in Fig. 4 for three different gate voltages. Note that the separation and the independent analysis of the two loss contributions represents an approximation and might thus lead to small deviations from the mathematically correct overall loss that can be obtained from the real part $\alpha$ of the complex propagation parameter $\underline{\gamma}$ according to Eq. (4).

For the contribution of the bare transmission line, we neglect $G'_{S,\text{bulk}}, G'_{S,\text{acc}}$ and $C'_S$ in Eq. (4), such that the loss can be written as

$$\alpha_{\text{TL}} = \Re\left\{\sqrt{(R'_{\text{TL}} + j\omega_{\text{RF}} L'_{\text{TL}})j\omega_{\text{RF}} C'_{\text{TL}}}\right\}. \tag{15}$$

For low frequencies $\omega_{\text{RF}} \ll \omega_{\text{TL}}$, $\omega_{\text{TL}} = R'_{\text{TL}}/L'_{\text{TL}}$, the loss $\alpha_{\text{TL}}$ increases in proportion to $\sqrt{\omega_{\text{RF}}}$,

$$\alpha_{\text{TL}} \approx \sqrt{\frac{R'_{\text{TL}} C'_{\text{TL}}}{2}} \sqrt{\omega_{\text{RF}}} \quad \text{for} \quad \omega_{\text{RF}} \ll \omega_{\text{TL}} \tag{16}$$

For higher frequencies, $\omega_{\text{RF}} \gg \omega_{\text{TL}}$, the loss $\alpha_{\text{TL}}$ of the bare transmission line approaches its low-loss, high-frequency approximation [38] and is constant with frequency,

$$\alpha_{\text{TL}} \approx \sqrt{C'_{\text{TL}}/L'_{\text{TL}}}\,(R'_{\text{TL}}/2) \quad \text{for} \quad \omega_{\text{RF}} \gg \omega_{\text{TL}}. \tag{17}$$

The frequency dependence of $\alpha_{\text{TL}}$ according to Eq. (15) is illustrated in Fig. 4 by the dotted black lines.

For the loss contribution of the RC-element that is formed by the slot capacitance $C'_S$ and the slab conductance $G'_S = G'_{S,\text{acc}} + G'_{S,\text{bulk}}$, we derive a relationship in Appendix C,

$$\alpha_{\text{RC}} = \frac{\omega_{\text{RF}}^2 C'^2_S G'^{-1}_S}{1 + (\omega_{\text{RF}} C'_S G'^{-1}_S)^2} \frac{1}{2\Re\left\{\underline{Z}_0^{-1}\right\}}. \tag{18}$$

For frequencies well below the RC corner frequency $\omega_{\text{RC}} = G'_S/C'_S$ of the RC element, we can approximate Eq. (18) by

$$\alpha_{\text{RC}} \approx \omega_{\text{RF}}^2 \frac{C'^2_S}{G'_S} \frac{1}{2\Re\left\{\underline{Z}_0^{-1}\right\}} \quad \text{for} \quad \omega_{\text{RF}} \ll \omega_{\text{RC}}, \tag{19}$$

leading to a quadratic increase of the losses with frequency. For frequencies well above the RC corner frequency $\omega_{\text{RC}}$, we obtain

$$\alpha_{\text{RC}} \approx G'_S \frac{1}{2\Re\left\{\underline{Z}_0^{-1}\right\}} \quad \text{for} \quad \omega_{\text{RF}} \gg \omega_{\text{RC}}, \tag{20}$$

i.e., the losses are independent of frequency. The frequency dependence of $\alpha_{\text{RC}}$ according to Eq. (18) is illustrated in Fig. 4 by the dashed black lines.

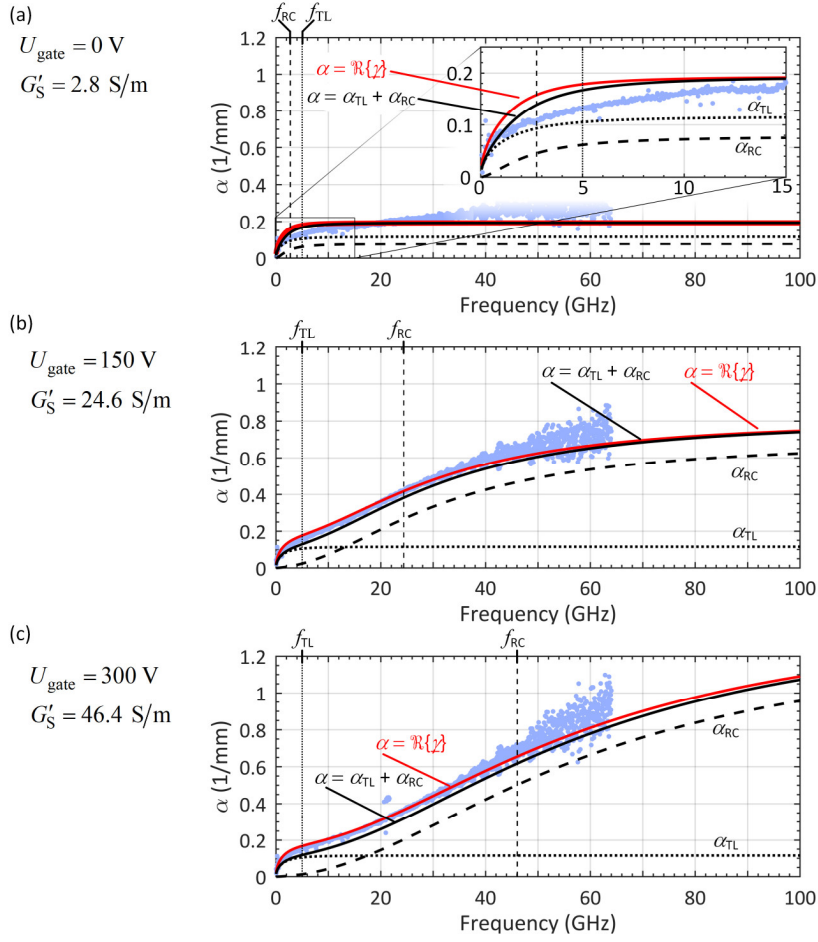

Fig. 4. Detailed representation of RF loss mechanisms in the slot-waveguide modulator. Blue data points represent measured data, and red traces correspond to the loss obtained by calculating the real part of the complex propagation parameter $\gamma$, Eq. (4), for the circuit parameters obtained by the fit. The three panels correspond to three values of the gate voltage $U_{\text{gate}}$ and represent slices of the 3D plot in Fig. 3(c). Besides the gate voltage $U_{\text{gate}}$, the associated slab conductance $G'_{\text{S}}$ is specified. The black dotted traces depict the loss contribution $\alpha_{\text{TL}}$ of the transmission line formed by the metal electrodes only, see Eq. (15), and the black dashed traces show the loss contribution $\alpha_{\text{RC}}$ of the RC element formed by the slab and the slot, see Eq. (18). The solid black line shows the sum of $\alpha_{\text{TL}}$ and $\alpha_{\text{RC}}$, which is in good agreement with the loss obtained from the full model represented by the red solid trace. **(a)** $U_{\text{gate}} = 0$ V: Except for very low frequencies, see inset, the loss is constant with frequency. **(b)** $U_{\text{gate}} = 150$ V: The quadratic frequency-dependence of $\alpha_{\text{RC}}$ for frequencies below the corner frequency $f_{\text{RC}} = \omega_{\text{RC}}/(2\pi)$ of the RC element, see Eq. (19), leads to a strong increase of the total loss with frequency. At frequencies $f_{\text{RF}} = \omega_{\text{RF}}/(2\pi) \gg f_{\text{RC}}$ the total loss approaches a constant value, see Eq. (17) and Eq. (20). **(c)** $U_{\text{gate}} = 300$ V: The corner frequency $f_{\text{RC}}$ is further increased compared to (b), and the total loss no longer reaches its constant high-frequency value in the considered frequency range.

The overall loss $\alpha$ can be estimated by adding $\alpha_{TL}$ and $\alpha_{RC}$, leading to the solid black lines in Fig. 4. These estimated losses are in good agreement with the loss obtained from directly using $\alpha = \Re\{\underline{\gamma}\}$ according to Eq. (4), see red lines in Fig. 4.

The three panels for Fig. 4(a, b, c) correspond to different gate voltages of 0 V, 150 V, and 300 V, respectively. For the samples used in our experiments, the doping of the slabs is rather low, and hence the resistance $G_S'^{-1}$ of the slabs is rather high in case no gate voltage is applied. This leads to a small corner frequency $\omega_{RC}$, illustrated by a dashed vertical line in Fig. 4(a), and the loss contribution $\alpha_{RC}$ of the RC-element quickly reaches the frequency-independent value given by Eq. (20). For low frequencies, the total loss is dominated by the contribution $\alpha_{TL}$ of the bare transmission line, Eq. (16), and increases in proportion to $\sqrt{\omega_{RF}}$. If $U_{gate}$ is increased, Fig. 4(b) and 4(c), $\omega_{RC}$ becomes larger and exceeds $\omega_{TL}$, which leads to a strongly frequency-dependent increase of the overall loss dominated by $\alpha_{RC} \sim \omega_{RF}^2$ for frequencies $\omega_{RF}$ between $\omega_{TL}$ and $\omega_{RC}$. At even higher frequencies ($\omega_{RF} \gg \omega_{RC}$), the total losses are again frequency-independent, see Eq. (17) and Eq. (20). For a gate voltage of 300 V, Fig. 4(c), the plateau, where the overall loss becomes frequency-independent, is no longer reached in the considered range of RF frequencies. For the discussion of design trade-offs (Section 5) it is important to note that in the 5…50 GHz range, increasing the slab conductivity, in the range considered here by changing $U_{gate}$, brings the modulator from a regime with weakly frequency-dependent loss to a regime with strongly frequency-dependent loss and may hence limit the bandwidth of the device.

## 4. Measured and modelled electro-optic response

With the knowledge of the six equivalent-circuit parameters $R_{TL}', L_{TL}', C_{TL}', G_{S,bulk}', G_{S,acc}', C_S'$ and of the optical group velocity $v_{g,opt}$ in the slot waveguide, Eq. (4 - 11) can be used to predict the EO frequency response of the SOH modulator. The optical group velocity $v_{g,opt} = c/n_{g,opt}$ is obtained by calculating the group refractive index $n_{g,opt} = 3.2$ by means of a numerical mode solver. In Fig. 5(a), the red lines show the EO frequency response as predicted by the model for a 750 μm-long modulator terminated with a 50 Ω impedance under the influence of gate voltages between 0 V to 300 V. These predictions are based on the circuit parameters listed in Table 1, as obtained from the purely electrical measurements of the frequency-dependent RF scattering parameters $\underline{S}_{21}$ and $\underline{S}_{11}$. Increasing the gate voltage increases the bandwidth, which is mainly due to a reduction of the RC time-constant of the slab-loaded slot waveguides. The predicted frequency response is compared against measured values obtained by using a VNA and a calibrated optical receiver [43], see blue traces in Fig. 5(a). In these measurements, the swept-source microwave stimulus of the VNA is applied to the modulator via microwave probes. The modulator is biased at its 3 dB point and the microwave signals amplitude is chosen small enough for the modulator to be operated under small-signal condition. The optical signal is subsequently amplified using an erbium-doped fiber amplifier (EDFA), filtered by a band-pass filter to remove out-of-band amplified-spontaneous-emission noise (ASE), and converted back to the electrical domain using a photodiode that is connected to the receiver port of the VNA. The frequency-dependent decay of the EOE frequency response $|\underline{m}_{EOE}(\omega_{RF})|$ according to Eq. (14) and Eq. (6) can be derived from the VNA measurement by the relation

$$10\log\left(|m_{EOE}(\omega_{RF})|^2\right) = S_{21,VNA,dB}(\omega_{RF}) - S_{21,VNA,dB}(\omega_{RF,0}). \quad (21)$$

In this relation, $S_{21,\text{VNA,dB}}$ denotes the scattering parameter measured by the VNA in dB after de-embedding from the frequency response of the photodiode and of the microwave probe, and $\omega_{\text{RF},0} = 2\pi \times 40$ MHz is a small reference frequency that is chosen as small as allowed by the VNA.

The different shades of blue in Fig. 5(a) indicate three independent measurements with small statistical variations. Overall, we find a good agreement of the measurements with the model predictions (red). In Fig. 5(b), both the 3 dB and the 6 dB EOE bandwidths of the modulator are shown for different gate voltages. For the 6 dB EOE bandwidth $f_{\text{6dB,EOE}}$, blue diamonds correspond to the measured values, and the dotted red line represents the bandwidths predicted by the model. Similarly, blue circles and the solid red line indicated measurements and model predictions of the 3 dB EOE bandwidth $f_{\text{3dB,EOE}}$. The good agreement of the predicted bandwidths with their measured counterparts validates the model introduced in Section 2, i.e., the proposed method allows to accurately predict the EOE frequency response of an SOH modulator based on the knowledge of purely electrical characteristics along with the group velocity $v_{\text{g,opt}}$ of the optical signal in the slot waveguide.

The electrical characteristics are easily accessible by either measurements in the RF domain, by simulations of the electrical RF device characteristics, or by estimating the distributed-element parameters from the modulator geometry and material properties. We believe that high-throughput wafer-level testing protocols might greatly benefit from a quantitatively verified model-based prediction of the EOE frequency response based on easily accessible electrical scattering parameters.

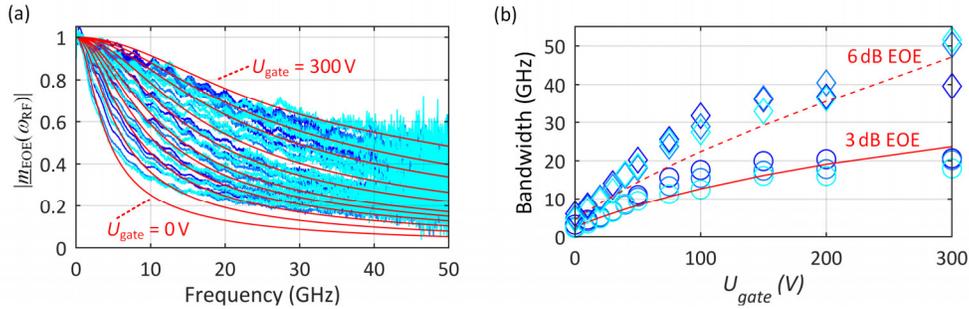

Fig. 5. Electro-optic-electric (EOE) frequency response $|m_{\text{EOE}}(\omega_{\text{RF}})|$ according to Eq. (6) and Eq. (14) of a 750 µm-long SOH modulator terminated with a 50 Ω impedance. **(a)** EOE response for different gate voltages $U_{\text{gate}}$ of 300 V, 200 V, 150 V, 100 V, 75 V, 50 V, 40 V, 30 V, 20 V, 10 V, and 0 V (from top to bottom). The red curve represents the prediction of the model according to Eq. (4) to Eq. (11), and the blue line corresponds to the measurement results, where the different shades indicate three independent measurements. With increasing gate voltage, the frequency response flattens. **(b)** 3dB EOE bandwidth $f_{\text{3dB,EOE}}$ (red solid line: model; blue circles: measurement) and 6 dB EOE bandwidth $f_{\text{6dB,EOE}}$ (red broken line: model; blue diamonds: measurement) as extracted from the curves in (a) as a function of applied gate voltage $U_{\text{gate}}$. The predicted bandwidths agree well with their measured counterparts, thus confirming that the proposed method allows to accurately predict the bandwidth of an SOH modulator based on the knowledge of purely electrical characteristics.

## 5. Design guidelines

### 5.1 Overcoming bandwidth limitations

Having validated the model of the SOH modulator, various design parameters can be tested for their impact on device performance. In the following, we assume a device with transmission line parameters $R'_{TL}, L'_{TL}$, and $C'_{TL}$ as specified in Table 1 and investigate the impact of variations of the slot capacitance $C'_S$ and of the slab conductance $G'_S$ on the EO bandwidth. This leads to the following considerations:

- The slab conductance $G'_S$ can be adjusted by modifying the doping profile in the slab. Increased slab conductivity will affect the electrical behavior and thus opens a way to increase the modulators bandwidth as discussed in detail below. As a trade-off, when increasing the slab conductivity by an increased doping concentration, the higher concentration of free carriers will also increase optical losses due to free carrier absorption (FCA).

- The slot capacitance $C'_S$ is mainly affected by the geometry of the slot waveguide and can be reduced by increasing the slot width. Reduced slot capacitance will affect the electrical behavior and thus the bandwidth as discussed in detail below. As a trade-off, increasing the slot width reduces the modulating electric RF field in the slot as well as the confinement factor of the optical field in the slot region. This can lead to a reduced modulation efficiency [23,44]. However, the poling efficiency may also be impaired by too narrow slot widths [45]. As an example, previous investigations of SOH modulators based on the EO material JRD1 [16] have shown that increasing the slot width to, e.g., 200 nm does not decrease the modulation efficiency of these devices.

The frequency response expressed by Eq. (6) considers all bandwidth-limiting effects, namely RC limitations of the slot waveguide together with the slabs, RF loss, impedance mismatch, and velocity mismatch. To investigate the impact of RC limitations, we first consider a theoretical modulator only limited by the RC-low-pass formed by the slab resistance and the slot capacitance, explicitly neglecting impedance mismatch, velocity mismatch and RF loss. The response of such a device is given by $\underline{H}_{RC}(\omega_{RF})$, Eq. (11). Figure 6 shows the 6 dB EOE bandwidth $f_{6dB,EOE}$ for a device that is only limited by the RC-low-pass as a function of slab conductance $G'_S$ and slot capacitance $C'_S$. The parameters of the devices used for experimental verification in this paper, see Table 1, are indicated by a green star for the case of 0 V gate voltage applied and by a magenta star for 300 V gate voltage applied. The bandwidth increases linearly with both $G'_S$ and $1/C'_S$. Note that the RC limitation is independent of the device length. Using the current devices as a reference, an increase of $G'_S$, e.g., by a factor of 20, possibly combined with a slight reduction of $C'_S$ by, e.g., 40 %, would permit RC-related EOE bandwidths $f_{\text{6dB,EOE}}$ in excess of 100 GHz if the device was only RC-limited. Increasing $G'_S$ by a factor of 20 can be realistically achieved without introducing excessive optical loss by using optimized doping profiles, see Section 5.2 for details.

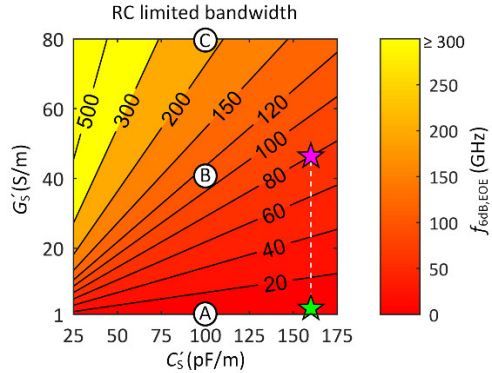

Fig. 6. 6 dB EOE bandwidth $f_{6dB,EOE}$ of a modulator only limited by the RC low-pass formed by the slot capacitance $C'_S$ and the slab conductance $G'_S$. Both a reduction of $C'_S$ and an increase of $G'_S$ linearly increases the bandwidth. The parameters of the devices used in this paper for experimental verification of the model, see Table 1, are indicated by a green star for the case of a gate voltage of 0 V and by a magenta star for a gate voltage of 300 V. For the parameters indicated by Ⓐ, Ⓑ, and Ⓒ, the frequency response of a 750 μm-long modulator is exemplarily shown in Fig. 8.

Next, we consider the impact of RF loss, impedance mismatch, and velocity mismatch on the modulator bandwidth. Figure 7(a) shows $f_{\text{6dB,EOE}}$ of a 750 µm-long modulator limited only by RF loss, impedance mismatch, and velocity mismatch, explicitly neglecting RC limitations. The response of such a theoretical device is given by Eq. (6) for $\underline{H}_{\text{RC}}(\omega_{\text{RF}}) = \text{const}$. For all slot capacitances $C'_S$, an increase in the slab conductance $G'_S$ tends to reduce the bandwidth, see Fig. 7(a). This can be understood by considering that increasing the slab conductance $G'_S$ initially brings the modulator from a regime with low and frequency-independent RF loss to a regime with high RF loss that increases with frequency, see Fig. 4. If the slab conductance $G'_S$ is increased further, the RF loss and its frequency dependence are reduced again, and the impact on the bandwidth decreases. In practical devices, however, $G'_S$ is constrained due to optical loss caused by FCA in the highly doped slab regions. This trade-off can be mitigated by optimized doping profiles, see Section 5.2.

We further investigate the overall RF loss $\alpha$ of the modulator, including both the loss in the metal electrodes and in the RC element formed by the slot waveguide and the slabs, as well as the associated real-part $Z_0$ of the characteristic impedance $\underline{Z}_0$ and the effective RF index $n_{\text{eff,RF}}$. The numerical values for $\alpha$, $Z_0$, and $n_{\text{eff,RF}}$ are shown exemplarily for an RF frequency of 50 GHz in Fig. 7(b), Fig. 7(c), and Fig 7(d), respectively. While the RF loss, Fig. 7(b), initially increases substantially with the slab conductance $G'_S$, the characteristic impedance $Z_0$, Fig. 7(c), stays close to 50 Ω. For the effective microwave index $n_{\text{eff,RF}}$, we observe a strong initial increase with the slab conductance $G'_S$, Fig. 7(d). This can be understood by considering that, for a low slab conductivity $G'_S$, the RC element is operated above its corner-frequency $\omega_{\text{RC}} < 50\,\text{GHz}$. In this regime, the slot capacitance $C'_S$ is only partially charged during one cycle. If the slab conductance $G'_S$ is increased, the slot capacitance becomes increasingly effective, and the RF wave is slowed down. For larger values of the slab conductance $G'_S$, the RC element is operated above its corner frequency and the capacitor is fully charged and de-charged during each cycle, leading to an RF index $n_{\text{eff,RF}}$ that is essentially independent of the slab conductance. Following the same argument, it can also be understood why an increase of the slot capacitance $C'_S$ therefore leads to an increase of $n_{\text{eff,RF}}$.

To investigate the frequency dependence of the RF loss, of the characteristic impedance, of the effective RF index, and of the RC low-pass as well as the associated impact on the modulator bandwidth, we consider three exemplary $C'_S$-$G'_S$-value-pairs for a 750 µm-long device. The chosen value pairs are marked by Ⓐ, Ⓑ, and Ⓒ in Fig. 6 and Fig. 7. The slot capacitance is fixed to $C'_S = 100\,\text{pF/m}$, while the slab conductance $G'_S$ is increased from $G'_S = 2.76\,\text{S/m}$ (Ⓐ), which corresponds to the value extracted from the device used in our experiment, to $G'_S = 40\,\text{S/m}$ (Ⓑ) and further to $G'_S = 80\,\text{S/m}$ (Ⓒ). Figure 8(a) shows, for the weakly conducting slab with $G'_S = 2.76$ S/m, the EOE frequency response $|m_{EOE}(\omega_{RF})|$, the RF loss $\alpha$, the characteristic impedance $Z_0$, and the effective RF refractive index $n_{eff,RF}$, all as a function of RF frequency up to 150 GHz. The solid red line depicts the total frequency response. Broken lines in blue, magenta, green, and black depict the frequency response by the sole contribution of the RC low-pass, the RF loss, the impedance mismatch, and the velocity mismatch, respectively. For a weakly conducting slab with $G'_S = 2.76$ S/m, the total bandwidth (red line) is limited almost exclusively by the RC low-pass (blue broken line), whereas the impact of RF loss (red magenta broken line), impedance-mismatch (green broken line) and of velocity mismatch (black broken line) are negligible, because $\alpha$, $Z_0$, and $n_{eff,RF}$ show little frequency dependence. If the slab conductivity is increased to, e.g., 40 S/m, Fig. 8(b), or 80 S/m, Fig. 8(c), the RC corner frequency increases. However, frequency-dependent RF loss starts to become significant and limits the overall bandwidth of the device to values smaller than the RC corner frequency. For the increased slab conductance $G'_S = 40\,\text{S/m}$ and $G'_S = 80\,\text{S/m}$, impedance mismatch and velocity mismatch do not significantly affect the bandwidth either. This confirms that RF loss is the most important bandwidth-limiting effect besides the RC time constant of the slab-slot configuration.

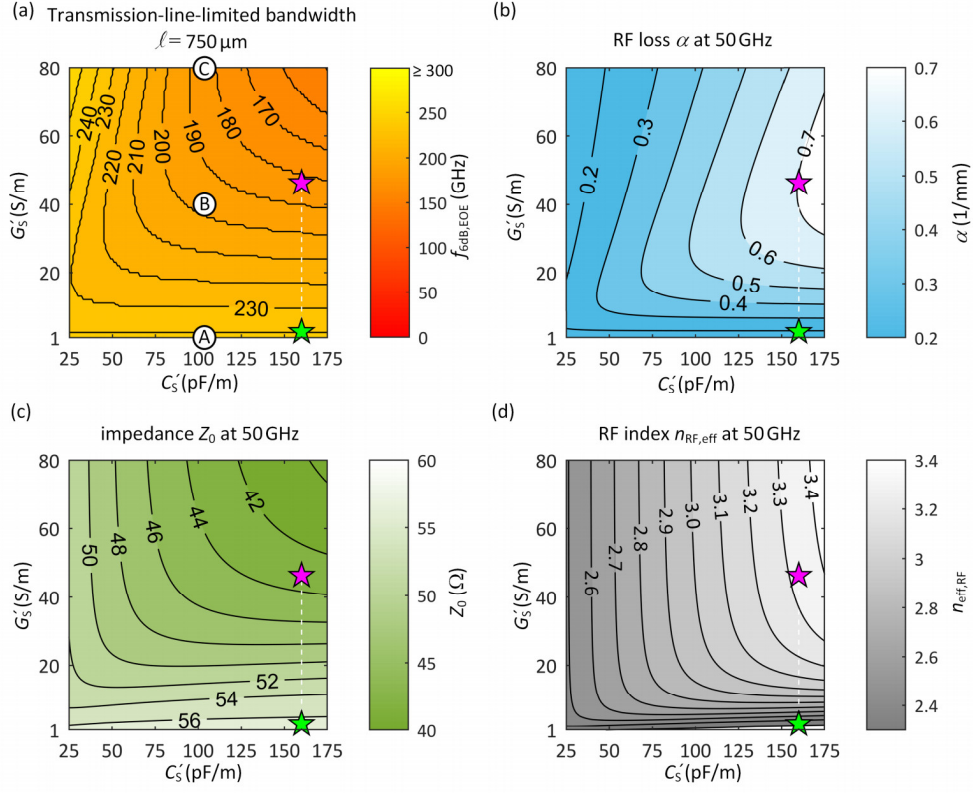

Fig. 7. **(a)** 6 dB EOE bandwidth $f_{6dB,EOE}$ of a 750 µm-long modulator limited only by RF loss, impedance mismatch, and velocity mismatch, explicitly neglecting the impact of the RC-low-pass formed by $C'_S$ and $G'_S$. For a large $C'_S$ and a low $G'_S$, an increase of $G'_S$ leads to a strong frequency-dependent increase of the RF loss and thus to a reduced bandwidth. For the parameters indicated by Ⓐ, Ⓑ, and Ⓒ the frequency response is exemplarily shown in Fig. 8. **(b)** RF loss for a 750 µm-long modulator at 50 GHz. The RF loss initially increases with increasing $G'_S$, see Fig. 4(a) and Fig. 4(b), but decreases again for very large values of $G'_S$, see Eq. (19). **(c)** Characteristic impedance $Z_0$ at 50 GHz. The impedance remains close to the system impedance of 50 Ω. **(d)** Effective RF refractive index $n_{eff,RF}$. For values of $G'_S \approx 50$ S/m and $C'_S \approx 130$ pF/m, $n_{eff,RF}$ is close to the effective optical group refractive index of 3.2 Increasing the slot capacitance slows down the RF wave and hence increases the RF refractive index. The detailed analysis in Fig. 8 shows that the bandwidth is not limited by velocity mismatch for practical values of $C'_S$ for the considered length of the modulator. The parameters of the devices (see Table 1) used in this paper for experimental verification of the model are indicated by a green star for the case of 0 V gate voltage applied and by a magenta star for 300 V gate voltage applied.

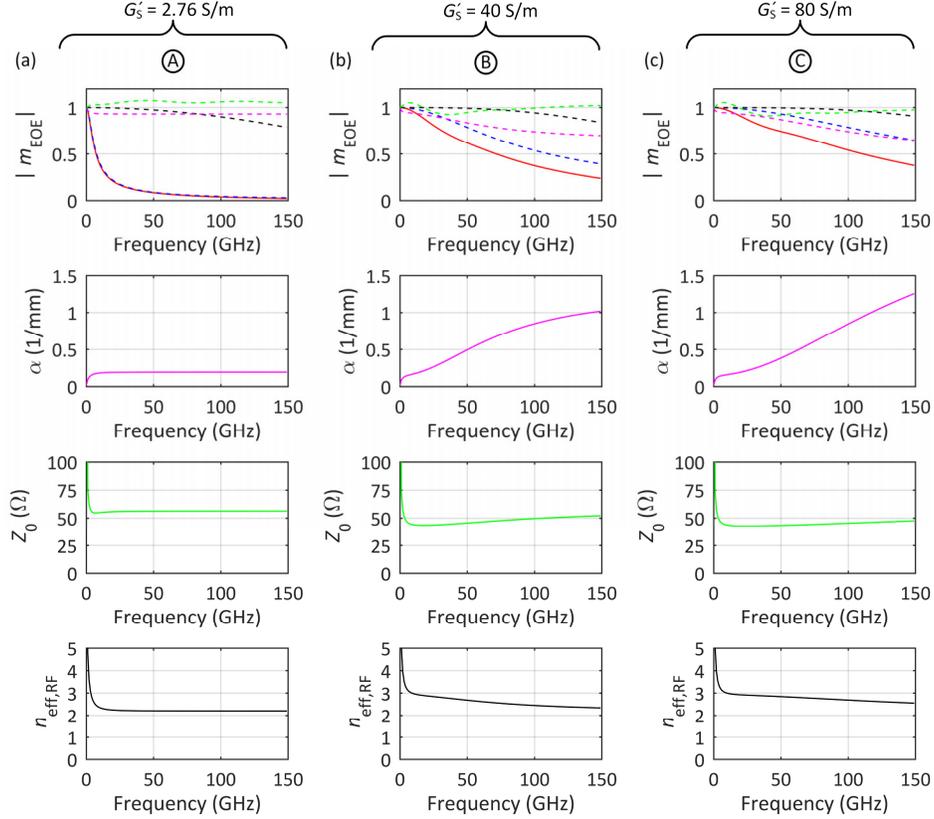

Fig. 8. Electro-optic bandwidth and RF properties for a 750 µm-long modulator with $C'_S = 100\,\text{pF/m}$ and **(a)** $G'_S = 2.76$ S/m, **(b)** $G'_S = 40$ S/m, and **(c)** $G'_S = 80$ S/m, corresponding to the points marked by Ⓐ, Ⓑ, and Ⓒ, respectively, in Fig. 6, Fig. 7(a), and Fig. 9(b). The first row shows the overall frequency response $|\underline{m}_{\text{EOE}}(\omega_{\text{RF}})|$ of the modulator (solid red line), as well as the frequency responses that correspond to the sole impact of RC limitations (blue dashed line), RF loss (magenta dashed line), impedance mismatch (green dashed line), and velocity mismatch (black dashed line). The second row shows the total RF loss α. In (a) the RF loss is low and frequency-independent, see Eq. (17). In (b) and (c) losses in the resistive slab are increased and the losses becomes strongly frequency-dependent, see Eq. (19), thereby reducing the bandwidth to values below the RC limit. The third row shows the characteristic impedance $Z_0$, which stays close to the system impedance of 50 Ω., except for low frequencies, where the $Z_0$ deviates from its ideal high frequency, low loss value, see Appendix B. The last row shows the effective RF refractive index $n_{\text{eff,RF}}$. In (a), the RC corner frequency $\omega_{\text{RC}}$ is small and the slot capacitance is not fully charged and de-charged as the RF wave propagates along the modulator. If the resistance of the slabs is reduced, see (b) and (c), $\omega_{\text{RC}}$ increases and the slot capacitance has a stronger impact on the RF wave, leading to a reduced velocity. The effective RF refractive index $n_{\text{eff,RF}}$ then becomes close to the optical group index $n_{\text{g,opt}} = 3.2$. Overall, impedance and velocity-mismatch are negligible for all of the three considered cases. In (a), the device is limited by the RC time constant of the slab-slot configuration. In (b) and (c) the RC time constant is reduced and the associated bandwidth is increased accordingly. At the same time, the increased RC corner frequency leads to an increased frequency-dependence of the RF loss, which causes additional bandwidth limitations.

Finally, we consider the 6 dB EOE bandwidth $f_{6\text{dB,EOE}}$ obtained from the full model as a function of the slab conductance $G'_S$ and the slot capacitance $C'_S$, taking into account all bandwidth-limiting effects, i.e., the RF loss, the impedance mismatch, the velocity mismatch and the RC low-pass. Figure 9 shows $f_{6\text{dB,EOE}}$ as obtained by the full model for four different lengths $\ell$ of the modulator. For long devices, the frequency dependence of the RF loss is more significant, and increasing the slab conductivity $G'_S$ has a smaller impact on the overall bandwidth than for short devices, see Fig. 9(a). Specifically, if $\ell$ is reduced to 500 µm, Fig. 9(c), and further to 200 µm, Fig. 9(d), the bandwidth approaches the RC-limited case shown in Fig. 6. A combination of reduced device length, reduced slab capacitance $C'_S$, and increased slab conductance $G'_S$ thus enables bandwidths well beyond 100 GHz.

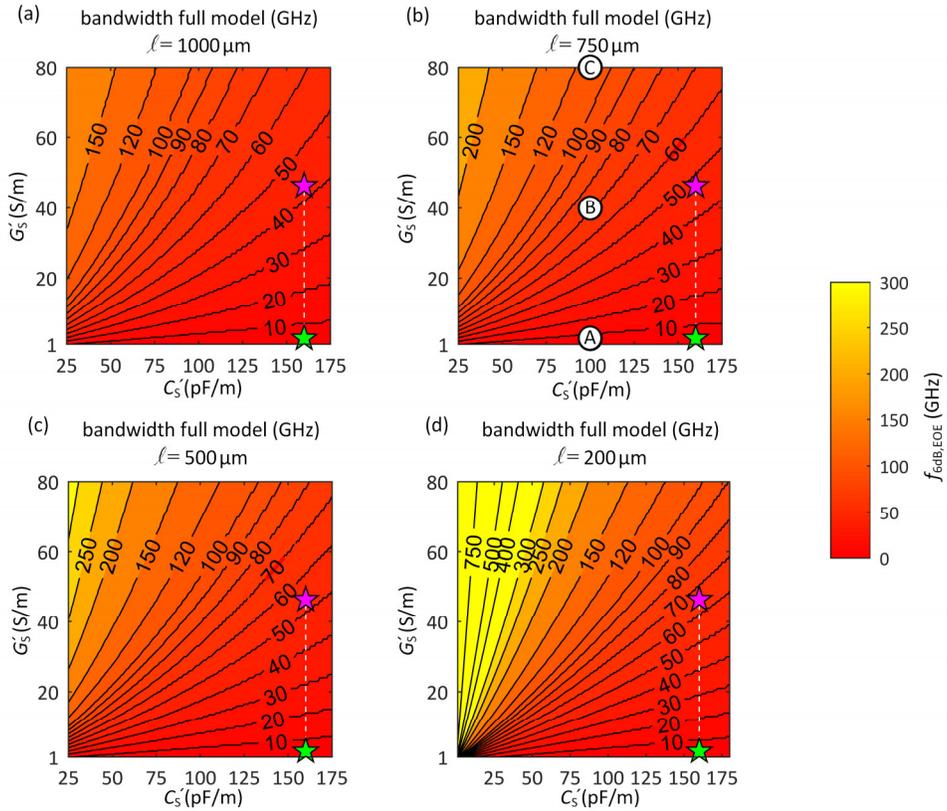

Fig. 9. 6 dB EOE bandwidth $f_{6\text{dB,EOE}}$ for modulators of different lengths $\ell$ based on the full model considering all bandwidth-limiting effects. The color scale is the same as in Fig. 6. If the length is reduced, the bandwidth of the modulator approaches the RC-limited case depicted in Fig. 6. The frequency response of the 750-µm-long modulator is shown in Fig. 8 for the $C'_S$-$G'_S$-values indicated by Ⓐ, Ⓑ, and Ⓒ in (b). The parameters of the devices used in this paper for experimental verification of the model, see Table 1, are indicated by a green star for the case of a gate voltage of 0 V and by a magenta star for a gate voltage of 300 V.

*5.2 Optimized doping profiles for low-loss modulators*

To increase the bandwidth of the modulators, highly conductive slab regions are desirable. Increasing the doping concentration near the optical slot waveguide, however, leads to optical loss due to free-carrier absorption (FCA). This effect can be mitigated by a two-step doping profile, see Fig. 1(b) [23,28,29]. Based on this approach, we suggest the following design procedure for SOH modulators:

(i) Given a certain EO material and a practical slot width, the targeted $\pi$-voltage $U_\pi$ will determine the length $\ell$ of the modulator.

(ii) Based on the slot width, which determines the slot capacitance $C'_S$, and the length $\ell$ of the modulator, the targeted bandwidth determines the slab conductance $G'_S$, see, e.g., Eq. 6 and Fig. 9.

(iii) For a given slab conductance $G'_S$, the optical loss can be minimized by optimizing the two-step doping profile [28,29]. This optimization is described in the following paragraph.

We exemplarily demonstrate the optimization of the doping profile for a device with a targeted slab conductance of $G'_S = 102$ S/m and a fixed slot width of 200 nm. For this device, a slot capacitance $C'_S$ of about 100 pF/m can be estimated by assuming a parallel-plate approximation. This assumption has to be taken with caution since fringing fields can contribute to a significant portion of the slot capacitance [23]. For the chosen values of $G'_S$ and $C'_S$, a device with length $\ell = 1$ mm, and transmission line parameters $R'_{TL}, L'_{TL}$, and $C'_{TL}$ as specified in Table 1, the model according to Eq. (6) would predict a bandwidth of $f_{6dB,EOE} = 114$ GHz. The geometry of the slot waveguide is fixed with a slab height of $h_{slab} = 70$ nm, a rail width $w_{rail} = 240$ nm, and a rail height of $h_{rail} = 240$ nm as shown in Fig. 10(a). The weakly doped region of the slab near the slot is depicted in green, and the more heavily doped region further away from the slot is depicted in blue in Fig. 10(a). As free parameters, we can vary the resistivity $\rho_{heavy}$ of the heavily doped region as well as the resistivity $\rho_{weak}$ and the width $w_{weak}$ of the weakly doped part.

For simplicity, we choose a fixed resistivity $\rho_{weak}$ of the weakly doped section in a first step, here $\rho_{weak} = 5.5 \times 10^{-4}$ Ωm. For a target conductance $G'_S = 102$ S/m of the slab, Fig. 10(b) shows the required bulk resistivity $\rho_{heavy}$ of the heavily doped region as a function of its width $w_{heavy}$. To reach the overall target conductance of $G'_S = 102$ S/m, the width $w_{weak}$ of the heavily doped region must stay below 1.16 µm. The bulk resistivity in the weakly and heavily doped regions can be translated into optical loss parameters using the experimentally obtained relations published in [46]. Figure 10(c) shows the bulk power absorption coefficient $\alpha_{Si}$ in the heavily and weakly doped silicon. Note that $\alpha_{Si}$ refers to the optical power – in contrast to the RF attenuation coefficient $\alpha$, specified in Eq. (4), which refers to the amplitude. The underlying model for the optical power absorption coefficient assumes a linear increase of the optical loss with the electrical conductivity in Si [46]. This assumption is only valid as long as the electrical conductivity is not limited by electron-impurity-scattering [47]. To translate the bulk absorption coefficient into a waveguide loss parameter, we compute the optical mode in the slot waveguide and the overlap with the doped silicon region [24] using a vectorial mode solver. The resulting power absorption coefficient $\alpha_{\mathrm{WG}}$ for the optical waveguide mode is shown in Fig. 10(d). If the weakly doped region is narrow, the heavily doped section comes close to the slot region leading to high optical loss. If the heavily doped section is far away from the slot waveguide, the conductivity has to be increased to maintain a given overall conductance $G'_{\mathrm{S}}$, which again increases the optical loss. A minimum excess loss of approximately 0.048 mm$^{-1}$ (0.2 dB/mm) is found for a weakly doped region with an intermediate width of $w_{\mathrm{weak}}$ = 1 µm and a medium resistivity of the heavily doped section of $\rho_{\mathrm{heavy}}$ = 1.43×10$^{-4}$ Ωm. Note that this doping-related excess loss does not significantly change the overall propagation loss of practically used slot waveguides, which is of the order of 0.5 dB/mm [48].

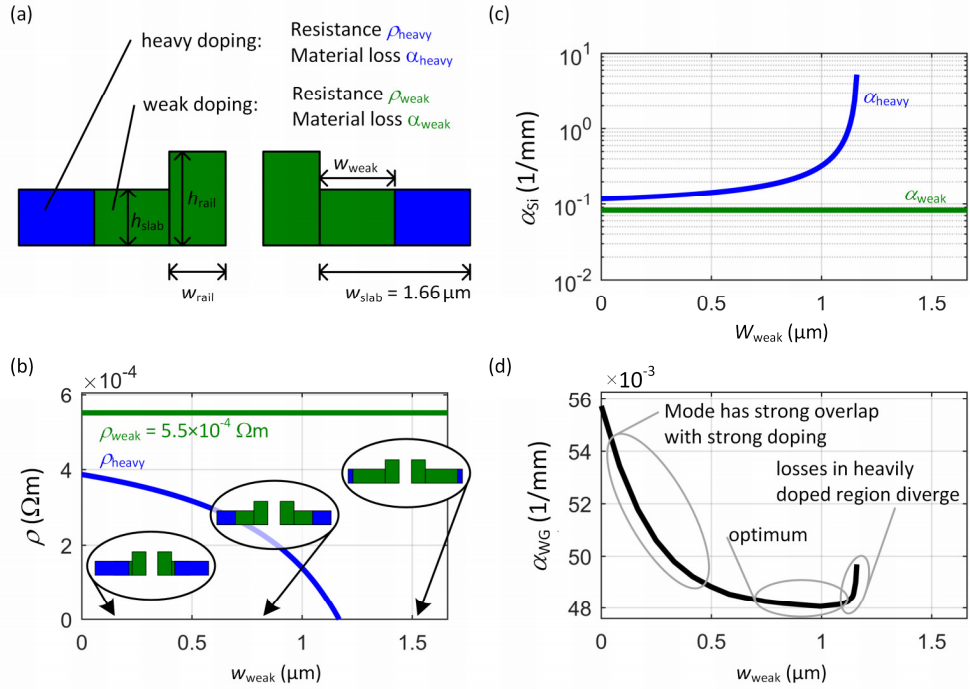

Fig. 10. Optimization of the doping profile in the slab-waveguide region to minimize optical loss. **(a)** Cross section of the slot waveguide and the adjacent slab: A low doping concentration is chosen for the rails of the slot waveguide and for the adjacent region of the slab (depicted in green), which interact strongly with the guided light. The doping is increased in the slab region that is further away from the rails (depicted in blue). We consider a waveguide with rails having a width of $w_{\mathrm{rail}}$ = 240 nm and a height of $h_{\mathrm{rail}}$ = 220 nm. The slabs have a height of $h_{\mathrm{slab}}$ = 70 nm and a width of $w_{\mathrm{slab}}$ = 1.66 µm. The width of the slot is 200 nm. For a fixed resistivity of the weakly doped silicon (here $\rho_{\mathrm{weak}}$ = 5.5×10$^{-4}$ Ωm), $w_{\mathrm{weak}}$ and $\rho_{\mathrm{heavy}}$ can be varied to realize the target conductivity of the slab (here $G'_{\mathrm{S}}$ =102 S/m). **(b)** Bulk resistivity $\rho_{\mathrm{heavy}}$ in the heavily doped slab regions that is required to reach the target conductivity, specified as a function of $w_{\mathrm{weak}}$ for fixed values of $\rho_{\mathrm{weak}}$ = 5.5×10$^{-4}$ Ωm and $w_{\mathrm{slab}}$ = 1.66 µm. **(c)** Corresponding material absorption coefficient in the doped silicon. Note that $\alpha_{\mathrm{Si}}$ refers to the power attenuation coefficient for the optical wave, whereas the symbol $\alpha$ in Eq. (4) refers to the amplitude attenuation coefficient of the corresponding RF wave. **(d)** Power loss coefficient experienced by the guided optical mode, obtained from the numerically calculated optical mode field and its overlap with the doped silicon regions. An optimum in terms of optical loss is found for an intermediate extension of the heavily doped section in combination with a medium resistivity.

So far, we have only considered a fixed resistivity $\rho_{\text{weak}} = 5.5 \times 10^{-4}$ Ωm of the weakly doped section. Varying this resistivity allows to further increase the bandwidth of the device or decrease its optical loss. This degree of freedom becomes particularly important when the length of the modulator is varied. For a quantitative analysis, we have repeated the optimization exemplarily described in the previous paragraph for devices of different lengths $\ell = 0.200$ mm, 0.5 mm, and 1.0 mm and for different resistivities $\rho_{\text{weak}} = 5.5 \times 10^{-4}$ Ωm, $16.5 \times 10^{-4}$ Ωm, and $49.5 \times 10^{-4}$ Ωm of the weakly doped region. Figure 11 shows the resulting optical excess loss caused by absorption in the doped silicon regions for modulators of different lengths $\ell$ as a function of their total bandwidth $f_{\text{6dB,EOE}}$. In this plot, the trade-off between the modulator figures of merits $U_\pi \propto \ell^{-1}$, bandwidth and insertion loss becomes evident. The numerical values presented in Fig. 11 suggest that it will be possible to realize SOH devices with a length of, e.g., 0.5 mm, having a bandwidth of $f_{\text{6dB,EOE}} = 100$ GHz while keeping the doping-related optical excess loss below 0.1 dB. Using highly efficient EO polymers [16], such a device would have a π-voltage of less than 1 V. The detailed device parameters leading to the results shown in Fig. 11 are depicted in Fig. 12. Figure 12(a) shows the relationship between the device bandwidth and the overall slab conductance $G'_S$ for the three different modulator lengths $\ell$ considered in Fig. 11. The graphs are obtained from the full model according to Eq. (6) in combination with a slot capacitance of $C'_S = 100$ pF/m and the transmission line parameters $R'_{\text{TL}}, L'_{\text{TL}}$, and $C'_{\text{TL}}$ as specified in Table 1. Figure 12(b) specifies the optimized resistivity $\rho_{\text{heavy}}$ of the heavily doped slab region along with the associated width $w_{\text{weak}}$ of the weakly doped region, see Fig. 12(c), and the associated power absorption coefficient $\alpha_{\text{WG}}$ for the optical waveguide mode, see Fig. 12(d), considering three fixed values of $\rho_{\text{weak}} = 5.5 \times 10^{-4}$ Ωm, $16.5 \times 10^{-4}$ Ωm, and $49.5 \times 10^{-4}$ Ωm.

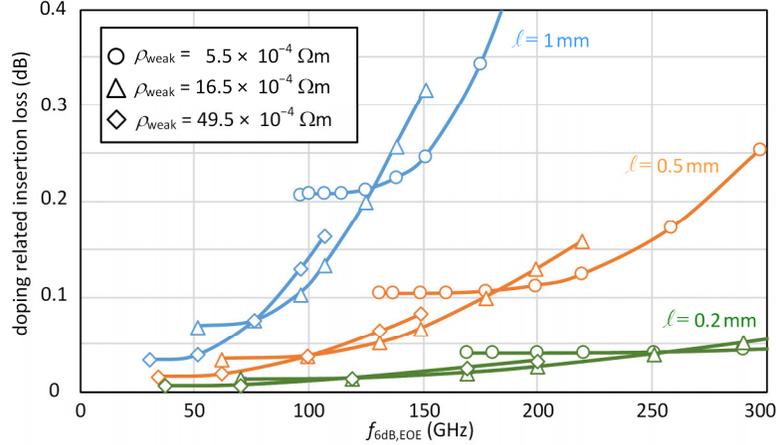

Fig. 11. Optical excess loss caused by absorption in the doped silicon regions for modulators of length $\ell = 0.2$ mm, 0.5 mm, and 1 mm as a function of the total 6 dB electro-optic-electric (EOE) bandwidth $f_{\text{6dB,EOE}}$. For each length, we consider three different resistivities $\rho_{\text{weak}}$ of $5.5 \times 10^{-4}$ Ωm, $16.5 \times 10^{-4}$ Ωm, and $49.5 \times 10^{-4}$ Ωm. The parameters $\rho_{\text{heavy}}$ and $w_{\text{weak}}$ are then optimized according to the procedure described in Fig. 9. For a given length of the modulator, there is a trade-off between electro-optic bandwidth and optical insertion loss. For a given bandwidth, optimizing the doping profile can lead to a reduction of optical loss. Data points are connected by interpolated lines as guide to the eye.

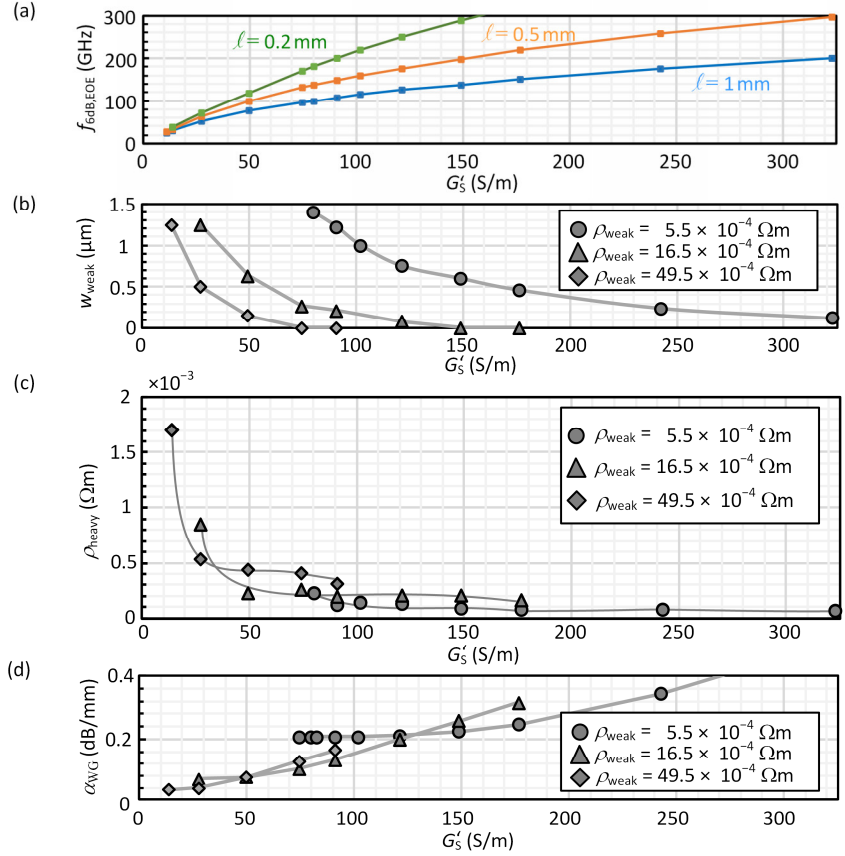

Fig. 12. Details of the optimized SOH modulator design for the modulators of length $\ell = 0.2$ mm, 0.5 mm, and 1 mm considered in Fig. 11. For each length, we consider three different resistivities $\rho_{\text{weak}}$ of $5.5 \times 10^{-4}$ Ωm, $16.5 \times 10^{-4}$ Ωm, and $49.5 \times 10^{-4}$ Ωm. The parameters $\rho_{\text{heavy}}$ and $w_{\text{weak}}$ are then optimized according to the procedure described in Fig. 9. **(a)** Relationship between device bandwidth $f_{\text{6dB,EOE}}$ and overall slab conductance $G'_S$ for three different modulator lengths. **(b)** Optimized width $w_{\text{weak}}$ of the weakly doped region, **(c)** the associated resistivity $\rho_{\text{heavy}}$ of the heavily doped slab region, and **(d)** the associated power absorption coefficient for the optical mode in the doped region of the silicon, all as a function of slab conductance $G'_S$, considering the three fixed values of $\rho_{\text{weak}}$. Data points in all subfigures are connected by interpolated lines as guide to the eye.

## 6. Summary

We formulate and experimentally validate an equivalent-circuit model of silicon-organic hybrid (SOH) slot-waveguide modulators based on distributed elements that describe the electric properties of the device. The model parameters can be extracted from purely electrical measurements, that provide the frequency-dependent scattering parameters of the travelling-wave device. With these parameters, we can accurately predict the small-signal electro-optic (EO) frequency response of the modulators, as confirmed by direct measurements of the EO frequency response. We further formulate design guidelines that lead to an optimum trade-off between EO bandwidth and optical insertion loss of SOH slot-waveguide modulators. We find that proper choice of doping concentrations and geometrical device parameters can enable SOH modulators with lengths of 0.5 mm and 6 dB electro-optic-electric (EOE) bandwidths of more than $f_{6dB,EOE} = 100$ GHz while maintaining π-voltages of less than 1 V and while keeping the optical excess loss caused by free-carrier absorption in the doped silicon regions below 0.1 dB.

**Appendix A: Mathematical relations for de-embedding the RF characteristics of on-chip modulators**

**Transfer matrices of contact pads**

The transfer matrix (T-matrix) of the left-hand-side contact pad including the subsequent transition to the modulator electrodes, see Fig. 2(a), is modelled by a series impedance $\underline{Z}$ and a shunt admittance $\underline{Y}$

$$\underline{T}_L = \begin{pmatrix} 1 & 0 \\ \underline{Y} & 1 \end{pmatrix} \begin{pmatrix} 1 & \underline{Z} \\ 0 & 1 \end{pmatrix} = \begin{pmatrix} 1 & \underline{Z} \\ \underline{Y} & \underline{Y}\underline{Z}+1 \end{pmatrix}. \tag{A1}$$

Similarly, the T-matrix of the right-hand-side contact pad including the subsequent transition to the modulator electrodes is modelled by the same series impedance $\underline{Z}$ and a the same shunt admittance $\underline{Y}$

$$\underline{T}_R = \begin{pmatrix} 1 & \underline{Z} \\ 0 & 1 \end{pmatrix} \begin{pmatrix} 1 & 0 \\ \underline{Y} & 1 \end{pmatrix} = \begin{pmatrix} \underline{Y}\underline{Z}+1 & \underline{Z} \\ \underline{Y} & 1 \end{pmatrix}. \tag{A2}$$

**Conversion between S-parameters and T-matrix parameters**

For conversion between S-parameters and electrical T-matrix parameters, we use the following relation [38,49]:

$$\underline{T}_{11} = \frac{1+\underline{S}_{11}-\underline{S}_{22}-\Delta\underline{S}}{2\underline{S}_{21}}, \tag{A3}$$

$$\underline{T}_{12} = \underline{Z}_{ref} \frac{1+\underline{S}_{11}+\underline{S}_{22}+\Delta\underline{S}}{2\underline{S}_{21}}, \tag{A4}$$

$$\underline{T}_{21} = \frac{1}{\underline{Z}_{ref}} \frac{1-\underline{S}_{11}-\underline{S}_{22}+\Delta\underline{S}}{2\underline{S}_{21}}, \tag{A5}$$

$$\underline{T}_{22} = \frac{1-\underline{S}_{11}+\underline{S}_{22}-\Delta\underline{S}}{2\underline{S}_{21}}, \tag{A6}$$

In these relations, $\Delta\underline{S} = \underline{S}_{11}\underline{S}_{22} - \underline{S}_{21}\underline{S}_{12}$, and $\underline{Z}_{ref}$ is the reference impedance of the measurement system.

**Transfer matrix of a transmission line**

The transfer matrix of a transmission line of length $\ell$ with propagation parameter $\underline{\gamma}$ and characteristic impedance $\underline{Z}_0$ is given by [38,49]

$$\underline{T} = \begin{pmatrix} \cosh(\underline{\gamma}\ell) & \underline{Z}_0 \sinh(\underline{\gamma}\ell) \\ \dfrac{1}{\underline{Z}_0}\sinh(\underline{\gamma}\ell) & \cosh(\underline{\gamma}\ell) \end{pmatrix}. \tag{A7}$$

**Appendix B: Low-frequency characteristic impedance of a transmission line**

The characteristic impedance given in its standard form by $\underline{Z}_0 = \sqrt{(R' + j\omega_{RF}L')/(G' + j\omega_{RF}C')}$ [38] is oftentimes considered for low-loss transmission lines at high frequencies only, where $\omega_{RF}L' \gg R'$ and $\omega_{RF}C' \gg G'$ holds. The characteristic impedance can then be approximated by a real-valued constant

$$Z_0 \approx \sqrt{L'/C'}. \tag{B1}$$

Note that this approximation is not valid for the low-frequency region where $\underline{Z}_0$ can be complex and frequency-dependent and the real and imaginary part may diverge for $\omega_{RF} \to 0$.

**Appendix C: Loss parameter of the RC element**

In this section we derive the RF amplitude attenuation parameter $\alpha_{RC}$ associated with the loss of the RC element formed by the slab resistance $G_S'^{-1}$ and the slot capacitance $C'_S$ following the approach in [23]. The active power dissipated in the RC element of an infinitesimally short section of the transmission line is linked to the real part of the complex line admittance $\underline{Y}'_{RC} \, dz$

$$dP = -\Re\left\{|\underline{U}|^2 \, \underline{Y}'^*_{RC}\right\} dz, \tag{C1}$$

where $\underline{U}$ denotes the voltage phasor and where the admittance $\underline{Y}'_{RC}$ of the RC element is given by

$$\underline{Y}'_{RC} = \left(\frac{1}{G'_S} + \frac{1}{j\omega_{RF}C'_S}\right)^{-1} = \frac{\omega_{RF}^2 C'^2_S G_S'^{-1} + j}{1 + \left(\omega_{RF} C'_S G_S'^{-1}\right)^2}. \tag{C2}$$

The active power that is transported on the entire transmission line is given by $P = \Re\left\{\dfrac{\underline{U}\underline{U}^*}{\underline{Z}_0^*}\right\}$, such that the ratio of the power lost in the RC element and the total power entering the infinitesimally short section is given by

$$\frac{dP}{P} = -\frac{\omega_{RF}^2 C'^2_S G_S'^{-1}}{1 + \left(\omega_{RF} C'_S G_S'^{-1}\right)^2} \frac{1}{\Re\left\{\underline{Z}_0^{-1}\right\}} dz. \tag{C3}$$

This differential equation is solved by $P = P_0 \exp(-2\alpha_{RC} z)$, where the additional factor of two in the argument arises from the fact that $\alpha_{RC}$ refers to the amplitude rather than the power attenuation. The amplitude attenuation parameter $\alpha_{RC}$ can thus be written as

$$\alpha_{RC} = \frac{\omega_{RF}^2 C'^2_S G_S'^{-1}}{1 + \left(\omega_{RF} C'_S G_S'^{-1}\right)^2} \frac{1}{2\Re\left\{\underline{Z}_0^{-1}\right\}}. \tag{C4}$$

**Funding**

(ERC) Consolidator Grant "TeraSHAPE" (773248); IARPA SuperCables Program ICENET (W911NF1920114), Alfried Krupp von Bohlen und Halbach Foundation; Helmholtz International Research School for Teratronics (HIRST); Karlsruhe School of Optics and Photonics (KSOP); Karlsruhe Nano-Micro Facility (KNMF)